\documentclass[twocolumn,showpacs,amsmath,nofootinbib,pra,aps,amssymb,longbibliography]{revtex4-1} %
\usepackage[T1]{fontenc}
\usepackage[utf8]{inputenc}
\usepackage{times}
\usepackage{color} 
\usepackage{array}
\usepackage{amssymb,amsmath}
\usepackage{amsbsy}
\usepackage{wasysym}
\usepackage[pdftex]{graphicx} 
\usepackage{bm}
\usepackage{float}
\usepackage{dcolumn}
\usepackage{siunitx}
\usepackage{scalerel}
\usepackage{tikz}
\usetikzlibrary{svg.path}

\definecolor{orcidlogocol}{HTML}{A6CE39}
\tikzset{
  orcidlogo/.pic={
    \fill[orcidlogocol] svg{M256,128c0,70.7-57.3,128-128,128C57.3,256,0,198.7,0,128C0,57.3,57.3,0,128,0C198.7,0,256,57.3,256,128z};
    \fill[white] svg{M86.3,186.2H70.9V79.1h15.4v48.4V186.2z}
                 svg{M108.9,79.1h41.6c39.6,0,57,28.3,57,53.6c0,27.5-21.5,53.6-56.8,53.6h-41.8V79.1z M124.3,172.4h24.5c34.9,0,42.9-26.5,42.9-39.7c0-21.5-13.7-39.7-43.7-39.7h-23.7V172.4z}
                 svg{M88.7,56.8c0,5.5-4.5,10.1-10.1,10.1c-5.6,0-10.1-4.6-10.1-10.1c0-5.6,4.5-10.1,10.1-10.1C84.2,46.7,88.7,51.3,88.7,56.8z};
  }
}

\newcommand\orcidicon[1]{\href{https://orcid.org/#1}{\mbox{\scalerel*{
\begin{tikzpicture}[yscale=-1,transform shape]
\pic{orcidlogo};
\end{tikzpicture}
}{|}}}}

\usepackage[unicode,breaklinks]{hyperref}
\hypersetup{
    unicode=true,
    plainpages=false, 
    colorlinks=true,
    linkcolor=blue,
    citecolor=blue,
    filecolor=black,
    urlcolor=blue
}
\AtBeginDocument{%
    \newwrite\bibnotes
    \def\bibnotesext{Notes.bib}
    \immediate\openout\bibnotes=\jobname\bibnotesext
    \immediate\write\bibnotes{@CONTROL{REVTEX41Control}}
    \immediate\write\bibnotes{@CONTROL{%
    apsrev41Control,author="08",editor="1",pages="1",title="0",year="1"}}
     \if@filesw
     \immediate\write\@auxout{\string\citation{apsrev41Control}}%
    \fi
}%

\newcommand{\gdd}{g^{\mathrm{dd}}}
\newcommand{\add}{a^{\mathrm{dd}}}
\newcommand{\fdd}{f^{\mathrm{dd}}}
\newcommand{\Phidd}{\Phi^{\mathrm{dd}}}
\newcommand{\edd}{\epsilon^{\mathrm{dd}}}
\newcommand{\eddeff}{\edd_\mathrm{eff}}
\newcommand{\bx}{\mathbf{x}}
\newcommand{\MM}{\mathcal{M}}
\newcommand{\ML}{\mathcal{L}}
\newcommand{\Ut}{\tilde{U}}
\renewcommand{\Re}{\operatorname{Re}}

\begin{document}

\title{Approximate theories for binary magnetic quantum droplets}

\author{Joseph~C.~Smith \orcidicon{0000-0002-2670-4044}}
\author{P.~B.~Blakie \orcidicon{0000-0003-4772-6514}}
\author{D.~Baillie \orcidicon{0000-0002-8194-7612}}

	\affiliation{Dodd-Walls Centre for Photonic and Quantum Technologies, New Zealand}
	\affiliation{Department of Physics, University of Otago, Dunedin 9016, New Zealand}
  
\date{\today}
\begin{abstract} 
We develop two approximate theories to describe the miscible and immiscible droplets that can occur in a binary mixture of highly magnetic bosonic atoms. In addition to allowing simpler calculations, the approximate theories provide insight into the role of quantum fluctuations in the two regimes. Results are validated by comparison to those from the extended Gross-Pitaevskii equation. As an application we solve for the ground state droplets crossing the miscible-immiscible transition as function of the short-ranged interspecies interaction parameter. We consider regimes where the transition occurs suddenly or as a smooth cross-over.  Using dynamical calculations we show that the character of the transition is revealed in the number of domains produced when ramping the droplet into the immiscible regime.
\end{abstract}

\maketitle

\section{Introduction}
Quantum droplets have been produced in ultra-dilute Bose gases in regimes where the meanfield driven collapse is arrested by a repulsive effective interaction arising from the quantum fluctuations. 
To date two classes of quantum droplet have been produced in experiments: i) Single component magnetic gases \cite{Ferrier-Barbut2016a,Chomaz2016a}, in which the atoms have long-ranged dipole-dipole interactions (DDIs) causing the droplets to take an elongated filament shape. ii) Binary gases of atoms with short-ranged interactions, including both mixtures of two spin components  \cite{Semeghini2018a,Cabrera2018a} and heteronuclear components  \cite{DErrico2019a,Guo2021a}.  These binary droplets \cite{Petrov2015a} occur in a miscible regime and have a spherical geometry.  

Recent experimental work has produced Bose condensates of different isotope mixtures of the highly magnetic atoms Er and Dy \cite{Trautmann2018a} and demonstrated magnetic Feshbach resonances for controlling their interspecies and intraspecies interactions \cite{Durastante2020a}. These developments motivated  two theoretical proposals for quantum droplet states in a binary magnetic gas (BMG) \cite{Smith2021a,Bisset2021a}: a system in which both the long-ranged DDIs and the rich interspecies interactions can occur. A key feature predicted for these BMG quantum droplets is the droplets can exist both in miscible and immiscible states, including regimes where one minority component can exist as an impurity trapped in the majority component.
The predictions presented in \cite{Smith2021a,Bisset2021a} were based on calculations used the extended Gross-Pitaevskii equation (EGPE), i.e.~a two-component Gross-Pitaevskii equation extended to include the leading order effect of quantum fluctuations. The expression for the quantum fluctuations  of the BMG requires numerical integration that depends upon the local value of the density of each component, making the general EGPE solution reasonably complicated to implement. Is is also possible to develop simpler descriptions based on a variational ansatz or the restrictive condition that both components have the same density profiles, but this limits the application to a small part of the miscible regime where both components have nearly identical wavefunctions. 

In this work we develop two approximate theories for quantum droplets of the BMG. The theories generally require the numerical solution of a Gross-Pitaevskii equation, but lead to a simpler analytic form for the quantum fluctuations. The first approximation is applicable to the immiscible regime, where one component dominates over the other at most locations in the droplet. The second approximation is applicable to the miscible regime, and is based on the assumption that both components have the same spatial mode, albeit with different relative amplitudes. By comparing the approximate solutions  to the EGPE solutions we identify the regimes where these approximations work well. Finally, we consider the miscible-immiscible ground state transition of BMG droplets controlled by varying the interspecies contact interaction. We demonstrate a case where the miscible-immiscible transition occurs suddenly and a case where the transition occurs via smooth cross-over, and show that the approximations usefully describe both cases. We use the dynamical EGPE theory to simulate the formation of domains after ramping across the transition to immiscibility.

This paper is organized as follows: In Sec.~\ref{s:form} we present a description of the energy due to quantum fluctuations in a homogeneous BMG and compute the resulting correction to the chemical potential. Then using the local density approximation we develop the EGPE description applicable to inhomogeneous droplets. 
We develop two simplified theories describing miscible and immiscible quantum droplets, the single mode approximation (SMA) and the immiscible approximation (IA) respectively. 
In Sec.~\ref{s:res} we numerically solve for ground droplet states of the system. In Sec.~\ref{s:SDS} we compare results from the EGPE to our simplified theories. In Sec.~\ref{s:imt} we solve for ground states across the miscible-immiscible transition, and finally in Sec.~\ref{s:dyn} we explore aspects of the miscible-immiscible transition using dynamic simulations of the time dependent EGPE. 

\section{Formalism}\label{s:form}

\subsection{EGPE formalism}
We consider a zero temperature mixture of two species of bosonic dipolar atoms.  We take both species to have equal mass $m$, which is a good approximation for any mixture of Er or Dy isotopes\footnote{The bosonic isotopes $^{162}$Dy,  $^{164}$Dy, $^{166}$Er, $^{168}$Er, and $^{170}$Er were condensed in    \cite{Trautmann2018a} with a  relative difference between mass extremes of less than $5\%$.}.
The total energy of the system is $E=\int d\bx \,\mathcal{E}$ where
\begin{align}
    \mathcal{E} &= -\frac{\hbar^2}{2m} \sum_i\psi_i^* \nabla^2 \psi_i  + \frac12 \sum_{i,j}n_i[g_{ij}n_j+\gdd_{ij}\Phidd_j  ]+ \mathcal{E}_\mathrm{QF},\label{e:Etotal}
\end{align}
is the energy density.
Here the atoms of component $i=1,2$ are described  by the macroscopic wavefunction $\psi_i$, with density  $n_i(\bx) = |\psi_i(\bx)|^2$. We normalize $\psi_i$ so that the total number is $N_i = \int d\bx\,n_i(\bx)$. 
We assume a fixed dipole moment $\mu_i^m$ polarized along the <C-D-X>$z$ direction. The dipole moments introduce a long-range anisotropic  DDI. 
The behavior of the DDI is described by 
\begin{align}
\Phidd_j(\bx) = \int d\bx'\fdd(\bx - \bx')n_j(\bx'),
\end{align} 
where 
\begin{align}
\fdd(\mathbf{r}) = \frac{3}{4\pi r^3}(1-3\cos^2\theta),
\end{align} is the DDI kernel.
The dipolar coupling constant is $\gdd_{ij} = 4\pi\hbar^2 \add_{ij}/m$ where $\add_{ij}=m\mu_0\mu_i^m\mu_j^m/12\pi\hbar^2$ is the dipole length. At short range the atoms interact by an $s$-wave contact interaction with coupling constant $g_{ij}=4\pi\hbar^2a_{ij}/m$, where $a_{ij}$ is the $s$-wave scattering length. Equation (\ref{e:Etotal}) goes beyond the standard meanfield theory by including the energy of the quantum fluctuations, described by the term
 $\mathcal{E}_\mathrm{QF}$, which we consider first in a homogeneous BMG. 

 \begin{figure}
    \includegraphics[width = 3.4in]{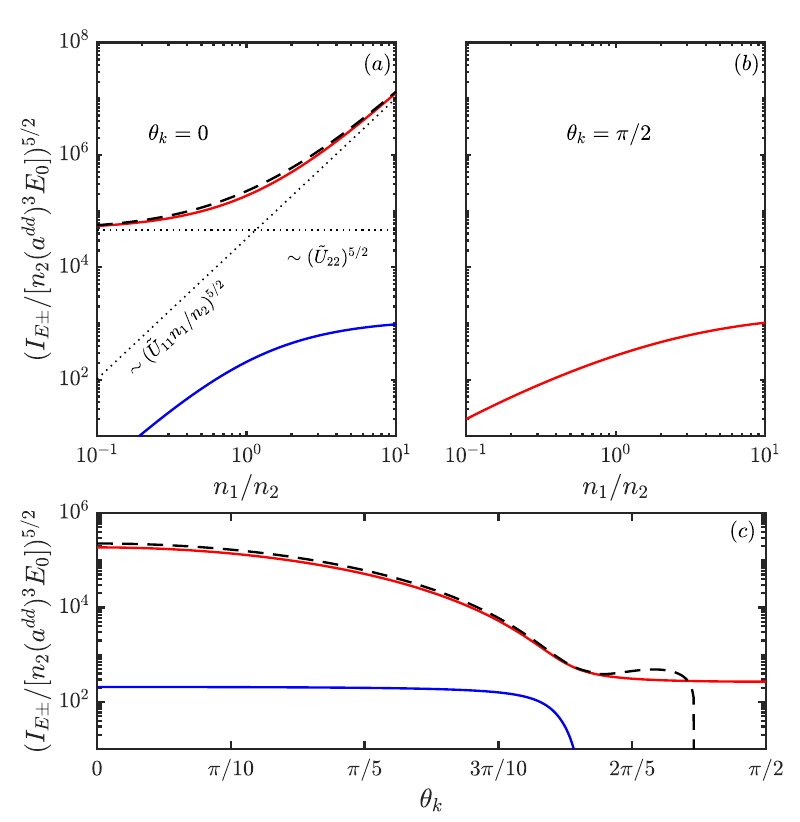}
    \caption{\label{f:LHY_integ} Real part of the quantum fluctuation energy integrand. $I_{E+}^{5/2}$ (solid red) and $I_{E-}^{5/2}$ (solid blue) from Eq.~\eqref{e:QF_integrand} and IA of $I_{E+}^{5/2}$ (dashed black) from Eq.~\eqref{e:IA_IEp} for $\{a_{11},a_{22},a_{12}\} = \{70,120,50\}a_0$ and, $\{\add_{11}$, $\add_{22}$, $\add_{12}\} = 130.1a_0 \equiv \add$ and  $E_0=\hbar^2/[m(\add)^2]$. In (a) $\theta_{k}=0$, (b) $\theta_{k} = \pi/2$ and the relative density ratio $n_{1}/n_{2}$ is varied. In (c)  $n_{1}/n_{2}=1$ and $\theta_{k}$ is varied.}
\end{figure}

\subsubsection{Quantum fluctuations of a homogeneous BMG}
The energy density of quantum fluctuations in a homogeneous BMG  with (constant) density $n_i$ in component $i$ is given by \cite{Bisset2021a}
\begin{equation}
\mathcal{E}_\mathrm{QF}^\text{hom}(n_1,n_2) = \frac{\sqrt2m^{3/2}}{15\pi^2\hbar^3} \sum_\pm \int_0^{\pi/2}d\theta_k \sin\theta_k I_{E\pm}^{5/2} \label{e:EQF},
\end{equation} 
where 
\begin{equation}
I_{E\pm}(\theta_k,n_1,n_2)  = n_1\Ut_{11} + n_2\Ut_{22} \pm\sqrt{\delta^2_1 + 4\Ut_{12}^2n_1n_2}\label{e:QF_integrand},
\end{equation}  
with $\delta_1 = n_1\Ut_{11} - n_2 \Ut_{22}$, and 
\begin{align}
\tilde{U}_{ij}=g_{ij}+\gdd_{ij}\left(\cos^2\theta_k-\frac{1}{3}\right),
\end{align}
being the Fourier transform of the total interaction potential $U_{ij}=g_{ij}\delta(\mathbf{r})+\gdd_{ij}\fdd(\mathbf{r})$.

We plot a selected case of the integrand $I_{E\pm}^{5/2}$ from Eq.~(\ref{e:EQF}) in Fig.~\ref{f:LHY_integ}. In general where $I_{E\pm}$ is significant we find that $I_{E+}$ is larger than $I_{E-}$. Also we observe the general scaling that $I_{E+}\sim \tilde{U}_{22}n_2$ when $n_1\ll n_2$,  and  $I_{E+}\sim \tilde{U}_{11}n_1$ when $n_2\ll n_1$ [see dotted lines on Fig.~\ref{f:LHY_integ}(a)]. In Fig.~\ref{f:LHY_integ}(b) we show $I_{E\pm}^{5/2}$ for $\theta_{k} = \pi/2$ seeing that at this angle $I_{E+}^{5/2}$ is much smaller compared to  $\theta_k=0$ ($I_{E-}^{5/2}$ is purely imaginary and is not shown).

In Fig.~\ref{f:LHY_integ}(c) we fix $n_{1}=n_{2}$ and examine the angular dependence of $I_{E\pm}^{5/2}$. As $\theta_k$ increases the two body interactions become less repulsive and $I_{E+}^{5/2}$ decreases. $I_{E-}^{5/2}$ is at first insensitive to changes in the angle however eventually decays to zero and then becomes purely imaginary for large enough $\theta_{k}$.

We also observe that, in the limit of vanishing DDIs, the result for the quantum fluctuations reduces to the simpler result established by Petrov   \cite{Petrov2015a}. In this regime $\Ut_{ij}\rightarrow g_{ij}$ and $I_{E\pm}$ becomes independent of $\theta_k$, yielding
\begin{equation}
\mathcal{E}_\mathrm{QF}^\text{hom} = \frac{\sqrt2m^{3/2}}{15\pi^2\hbar^3} \sum_\pm \Bigl(n_1g_{11} + n_2 g_{22} \pm\sqrt{\delta^2_1 + 4 g_{12}^2n_1n_2}\Bigr)^{5/2}\!\!.
\end{equation}
This expression can be written as,
\begin{equation}
\mathcal{E}_\mathrm{QF}^\text{hom} = \frac{8m^{3/2}}{15\pi^2\hbar^3} \left(n_1g_{11}\right)^{5/2}f\left(1,\frac{g^{2}_{12}}{g_{11}g_{22}},\frac{n_{2}g_{22}}{n_{1}g_{11}}\right),
\end{equation}
where $f(1,x,y) = \sum_{\pm} (1+y\pm\sqrt{(1-y)^2 + 4xy})^{5/2}/4\sqrt{2}$.

\subsubsection{Quantum fluctuations in the local density approximation}
To apply result (\ref{e:EQF}) to the inhomogeneous system we set the quantum fluctuation energy density in Eq.~(\ref{e:Etotal}) to
\begin{align}
 \mathcal{E}_\mathrm{QF}\to \Re\{\mathcal{E}^\text{hom}_\mathrm{QF}[n_1(\bx),n_2(\bx)]\}.\label{EQFreal}
\end{align}

The imaginary component of the homogeneous quantum fluctuation energy occurs due to a phonon instability in modes with wavelength exceeding the droplet size, which is typically of the order of $1\mu$m (see \cite{Bisset2016a}). Thus, for the inhomogeneous system, we only include the real part of the quantum fluctuations. 

In the non-dipolar  mixtures, alternatives to this approximation have been explored. This includes the use of many-body density pairing theory \cite{liu2020a}, and diffusion Monte Carlo methods  \cite{Astrakharchik2020a}. In general these methods are in qualitatively good agreement with the theory  in \cite{Petrov2015a}.





\subsubsection{Extended Gross-Pitaevskii equation}

We are interested in stationary states of Eq.~(\ref{e:Etotal}). We compute $\delta E / \delta \psi^{*}_{i}$ yielding two coupled EGPEs
\begin{align}
\ML_i\psi_i = \mu_i\psi_i,\label{EGPE}
\end{align} 
where $\mu_i$ is the chemical potential of the $i$-th component and 
\begin{equation}
    \ML_i = -\frac{\hbar^2\nabla^2}{2m} + \sum_j[g_{ij}n_j(\bx)+\gdd_{ij}\Phidd_j(\bx)] + \Delta\mu_i\label{e:GP_operator},
\end{equation}
where $\Delta\mu_i$ is the correction to the chemical potential resulting from quantum fluctuations and is readily calculated from Eq.~\eqref{EQFreal}. For the $i$-th component $\Delta\mu_i = \partial_{n_i}\mathcal{E}_\mathrm{QF}$ giving
\begin{equation}
    \Delta\mu_1= \frac{m^{3/2}}{3\sqrt2\pi^2\hbar^3}\sum_\pm\int_0^{\pi/2} d\theta_k\sin\theta_k \Re I_{1\pm}\label{e:Deltamu},
\end{equation}
where 
\begin{equation}
    I_{1\pm} = \left( \Ut_{11} \pm \frac{\delta_1\Ut_{11}+2\Ut_{12}^2n_2}{\sqrt{\delta_1^2+4\Ut_{12}^2n_1n_2}}\right) I_{E\pm}^{3/2}.
\end{equation}
A similar calculation with $\delta_2=-\delta_1$ gives an expression for $\Delta\mu_2$ which we omit here.

\subsection{Immiscible approximation (IA)}

When the mixture is immiscible the two components will phase-separate and domains form. Each domain is dominated by a single component and separated by narrow interface region where the two components have equal density.
%
%
The mixture can then be well represented by a small $n_2$ expansion for Eq.~\eqref{e:QF_integrand} in the component 1 domain and a small $n_1$ expansion in the component  2 domain. We add these two expansions together to give the IA approximation to  \eqref{e:QF_integrand}
\begin{align}
    I_{E+}^{5/2}  &= (2n_1\Ut_{11})^{5/2} + 5(2n_1)^{3/2}n_2\Ut_{11}^{1/2}\Ut_{12}^2 + O(n_2^2n_1^{1/2})\nonumber\\
&+(2n_2\Ut_{22})^{5/2} + 5(2n_2)^{3/2}n_1\Ut_{22}^{1/2}\Ut_{12}^2 + O(n_1^2n_2^{1/2}),\label{e:IA_IEp}
\end{align}
%
%
where the first line is valid for small $n_2$, and the second line is valid for small $n_1$. When the expansion for one component is invalid, those results are dominated by the valid expansion. For example, in the small $n_{1}$ expansion [second line of Eq.~\eqref{e:IA_IEp}] the second term is of order $n_{1}$ and is invalid for large $n_{1}$, however the small $n_{2}$ expansion  [first line of Eq.~\eqref{e:IA_IEp}] has two terms of order $ n_1^{5/2}$ and $n_1^{3/2}$ which dominate over this term. We have not included the expansion of $I_{E-}$ as it is higher order. 

We have plotted Eq.~\eqref{e:IA_IEp} in Fig.~\ref{f:LHY_integ} and find for $\theta_{k}=0$ [Fig.~\ref{f:LHY_integ}(a)] the expansion works exceptionally well when either $n_{1}\ll n_{2}$ or $n_{2} \ll n_{1}$.  When $n_{1} \approx n_{2}$ (the worst case for this approximation), we find the IA slightly over estimating the actual value of Eq.~\eqref{e:QF_integrand}, but it still qualitatively describes the behavior when $I_{E+}$ is significant [Fig.~\ref{f:LHY_integ}(c)]. 

In order to compute the IA chemical potential correction to the $i$-th component we differentiate Eq.~\eqref{e:IA_IEp} with respect to $n_i$ to obtain
\begin{align}
I_{1+} &\approx 2^{5/2}  \left[ n_1^{3/2} \Ut_{11}^{5/2} + \left( \tfrac32 \sqrt{n_1} n_2 \Ut_{11}^{1/2} + n_2^{3/2} \Ut_{22}^{1/2}\right)\Ut_{12}^2\right],\\ 
I_{2+} &\approx 2^{5/2}  \left[ n_2^{3/2} \Ut_{22}^{5/2} + \left( \tfrac32 \sqrt{n_2} n_1 \Ut_{22}^{1/2} + n_1^{3/2} \Ut_{11}^{1/2}\right)\Ut_{12}^2\right].
\end{align}


Then $\Delta\mu_1$ can be written as
\begin{align}
\Delta\mu_1 &\approx \frac{4m^{3/2}}{3\pi^2\hbar^3}\Bigl[n_1^{3/2}(\gdd_{11})^{5/2}J(1/\edd_{11},1/\edd_{11})\nonumber\\
&+\frac32\sqrt{n_1}n_2(\gdd_{12})^2(\gdd_{11})^{1/2}J(1/\edd_{11},1/\edd_{12})\nonumber\\
&+ n_2^{3/2}(\gdd_{12})^2(\gdd_{22})^{1/2}J(1/\edd_{22},1/\edd_{12})\Bigr]\label{e:IAcorrection},
\end{align}
where we define the dimensionless quantity $\edd_{ij} = \gdd_{ij}/g_{ij}$ and the integral
\begin{equation}
    J(x,y) = \Re \int_0^1 du\, (x+3u^2 - 1)^{1/2} (y+3u^2 - 1)^2\label{e:IA_Jxy}.
\end{equation}
For $x\ge-2$, $J(x,y)=0$, for $x\ge1$ the integral is real, and otherwise the real part is found with the lower limit $\sqrt{(1-x)/3}$.

%
%
We thus obtain what we refer to as the EGPE in the IA
\begin {align}
\mathcal{L}_i^\text{IA}\psi_i=\mu_i\psi_i,\label{EGPEIA}
\end{align}
where $\mathcal{L}_i^\text{IA}$ is as in Eq.~(\ref{e:GP_operator}) but with $\Delta \mu_i$ evaluated according to Eq.~(\ref{e:IAcorrection}). This has the advantage that we can evaluate Eq.~\eqref{e:IA_Jxy}  analytically as
\begin{align}
J(x,y) = p_1(x,y) \sqrt{x+2} +p_2(x,y)\ln\frac{\sqrt{x+2} + \sqrt3}{\sqrt{|x-1|}}, \label{e:IA_J_eval}
\end{align}
for $x\ge-2$ and zero otherwise (with the second term zero for $x=1$). We have introduced polynomials
\begin{align}
    p_1(x,y) &= \frac1{16}(-x^2+4xy+8y^2+4y+9)\label{e:IA_P1},\\
p_2(x,y) &= \frac{x-1}{16\sqrt3}(x^2-4xy+2x+8y^2-12y+5)\label{e:IA_P2}.
\end{align}

When $n_2\approx0$, Eq.~\eqref{e:IAcorrection} reduces to the single component limit 
\begin{align}
    \Delta\mu_1 &\approx \frac{4m^{3/2}}{3\pi^2\hbar^3}n_1^{3/2}(\gdd_{11})^{5/2}\MM_5(1/\edd_{11}), \label{e:singlemu}
\end{align}
where $\MM_5(x) = J(x,x)$ \cite{Smith2021a}, and when $n_1\approx0$, Eq.~\eqref{e:IAcorrection} reduces the impurity limit found in Ref.~\cite{Bisset2021a} which we can evaluate analytically as
\begin{equation}
\Delta\mu_1\approx \frac{4m^{3/2}}{3\pi^2\hbar^3}n_2^{3/2}(\gdd_{12})^2(\gdd_{22})^{1/2} J(1/\edd_{22},1/\edd_{12})\label{e:IA_Deltamu}.
\end{equation}

\subsection{Single mode approximation}\label{s:sma}
\subsubsection{Single modal $\psi_i$}\label{sec:singlemodal}
The meanfield effective potential of the GPE for a miscible BMG, $\sum_j[g_{ij}n_j(\bx)+\gdd_{ij}\Phidd_j(\bx)]$ is the same for both components  
if $\mu_{1}^{m} = \mu_{2}^{m}$, $\psi_1/\sqrt{N_1} = \psi_2/\sqrt{N_2}$, and  
\begin{align}
\frac{N_{1}}{N_{2}}  &= \frac{g_{22}-g_{12}}{g_{11}-g_{12}}.\label{smaCrit}
\end{align}
The solutions of the GPE can then be single modal, both equal to $\psi(\bx) =\psi_i(\bx)\sqrt{N/N_i}$ where $N=N_1+N_2$ is the total number of atoms.
\subsubsection{Single mode approximation}

We wish to extend the notion of single-modality to the EGPE in what we call the single mode approximation (SMA) and write $\psi_i(\bx)= \psi(\bx)\sqrt{N_i/N}$.
%
We substitute our expression for $\psi_{i}$ into Eq.~\eqref{EQFreal} and compute $\delta E/\delta \psi^{*}$ . This leads to a single-component effective equation given by 
\begin{equation}
	\ML^{\text{sma}}\psi=\mu\psi,
\end{equation} 
with
\begin{align}
    \ML^{\text{sma}}\!&=\!-\frac{\hbar^2\nabla^2}{2m} \!+\! g_\mathrm{sma}n(\bx) \!+\! \gdd_\mathrm{sma}\Phidd_\mathrm{sma}(\bx) + \gamma_\mathrm{sma}[n(\bx)]^{3/2}\!,\label{e:Lsma}
\end{align}
where  $g_\mathrm{sma} = \sum_{i,j} g_{ij} N_i N_j/N^2$, and $\gdd_\mathrm{sma} = \sum_{i,j} \gdd_{ij} N_i N_j/N^2$ 
are the effective $s$-wave and dipolar coupling constants, respectively. The effective dipolar interactions are described by $\Phidd_\mathrm{sma}(\bx) = \int d\bx' \fdd(\bx-\bx')n(\bx'),$ where $n(\bx)=|\psi(\bx)|^2$. The coefficient of the quantum fluctuation term is given by
\begin{align}
    \gamma_\mathrm{sma} &= \frac{ m^{3/2}}{3\sqrt2\pi^2\hbar^3 N^{5/2}}  \sum_\pm\int_0^{\pi/2} d\theta_k \sin\theta_k \Re\: (I_\mathrm{sma}^\pm)^{5/2}\label{e:SMAgamma},
\end{align}
where
\begin{align}
    I_\mathrm{sma}^{\pm}\! &=\!  N_1 \Ut_{11}\!+\!N_2 \Ut_{22}\!\pm\!\sqrt{(N_1 \Ut_{11}\!-\!N_2\Ut_{22})^2\!+\!4\Ut_{12}^2N_1N_2}. \label{e:Isma}
\end{align}

Notice that $\gamma_\mathrm{sma}$ has no dependence on the density, so only needs to be computed once for a given set of parameters.

\subsubsection{Reduction to same shape approximation}
We now wish to shown how our more general single mode approximation connects to the same shape approximation used in \cite{Smith2021a}. First consider the case where $N_1=N_2$, $g_{11}=g_{22}$, and $\mu_i^m = \mu_j^m$, then Eq.~\eqref{e:SMAgamma} simplifies to
\begin{align}
    \gamma_\mathrm{sma} &= \frac{m^{3/2}}{3\sqrt2\pi^2\hbar^3} [\max(0,g_{ii}-g_{12})^{5/2} \notag\\& + (2\gdd_{ii})^{5/2} \MM_5(1/\eddeff)], \label{e:SMAgammaEval}
\end{align}
with $\eddeff = 2\gdd_{ii}/(g_{ii}+g_{12})$. The coupling constants reduce to  $g_\mathrm{sma}=(g_{ii}+g_{12})/2$ and $\gdd_\mathrm{sma} = \gdd_{ii}$. Substituting $n(\bx) = 2n_i(\bx)$ in Eq.~\eqref{e:Lsma} gives the same shape approximation of \cite{Smith2021a}.

\subsubsection{Non-magnetic binary quantum droplets}
Finally it is worth briefly discussing the connection of the SMA to a widely used approach to describe quantum droplets occurring in binary mixtures without DDIs, as first developed by Petrov \cite{Petrov2015a}. 
These droplets emerge in a particular parameter regime $0<-\delta g\ll g_{11},g_{22}$, which ensures that the two components are miscible. Under these conditions the densities of the two components are essentially locked to a ratio to ensure an energy minimising attractive effective interaction parameter of  $\delta g=g_{12}+\sqrt{g_{11}g_{22}}$. In this situation the two components exhibit single modality with $N_1/N_2=\sqrt{g_{22}/g_{11}}$, which is equivalent to Eq.~(\ref{smaCrit}) to leading order in $\delta g$.

\section{Results}\label{s:res}
\subsection{Computational methods}

We solve the stationary states of the EGPE, SMA, and IA using gradient flow \cite{Bao2010a,Lee2021a}, also known as the imaginary time method. We discretize the time-derivative in the gradient flow equations using a backward-forward Euler scheme, where we treat the kinetic energy using a backward Euler step and the other terms using a forward Euler step (see \cite{Lee2021a}).

We solve for ground states with cylindrical symmetry and choose a Fourier-Bessel numerical representation in which our system is discretized in the radial direction using a Bessel grid and in the axial direction using a Fourier grid. 
We find that when solving for an immiscible state, the interface between the two components can become quite sharp, requiring many grid points to resolve accurately. We evaluate the DDI using the convolution theorem, in which the cylindrically cut-off DDI kernel in \cite{Lu2010a} is used.

When calculating the stationary states of Eq.~\eqref{EGPE}, we need the chemical potential corrections $\Delta \mu_i$ at all points of space at each iteration.  While  $\Delta \mu_i$ explicitly depends on both $n_1$ and $n_2$, we can simplify this to  a form more suitable for numerical calculations. To do this we write the chemical potential corrections in terms of the ratio $n_{1}/n_{2}$ [cf.  Eq.~\eqref{e:Deltamu}]
\begin{equation}
    \frac{\Delta\mu_i}{n_2^{3/2}} = \frac{m^{3/2}}{3\sqrt2\pi^2\hbar^3}\sum_\pm\int_0^{\pi/2} d\theta_k\sin\theta_k \Re I^\uparrow_{i\pm}(n_1/n_2)\label{e:Deltamu_scaled},
\end{equation}
where  
\begin{align}
    I^\uparrow_{1\pm}\left(n_1/n_2\right) &= \left( \Ut_{11} \pm \frac{\delta^\uparrow_1\Ut_{11}+2\Ut_{12}^2}{\sqrt{(\delta^\uparrow_1)^2+4\frac{n_1}{n_2}\Ut_{12}^2}} \right) (I^\uparrow_{E\pm})^{3/2},\label{eq:Eup} \\
 \delta^\uparrow_1\left(n_1/n_2\right) &= \frac{n_1}{n_2} \Ut_{11} - \Ut_{22},\\
I^\uparrow_{E\pm}\left(n_1/n_2\right) &= \frac{n_{1}}{n_{2}}\Ut_{11} + \Ut_{22} \pm\sqrt{(\delta^\uparrow_1)^2 + 4\frac{n_1}{n_2}\Ut_{12}^2}\label{e:QF_integrand_scaled}.
\end{align}  
The expression for $I^\uparrow_{2\pm}$ takes a similar form.

\begin{figure}
    \includegraphics[width = 3.4in]{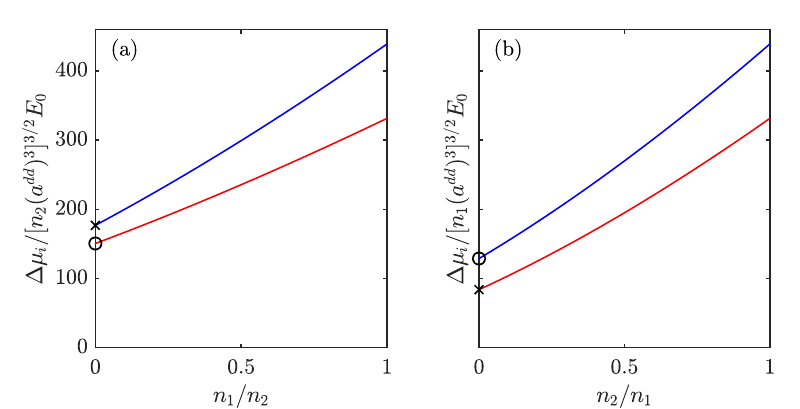}
    \caption{\label{f:f1D_lookup}  (a) Chemical potential correction $\Delta\mu_1/n_2^{3/2}$ (red) and $\Delta\mu_2/n_2^{3/2}$ (blue) from \eqref{e:Deltamu_scaled} in terms of the ratio $n_1/n_2$ and, (b) similarly $\Delta\mu_i/n_1^{3/2}$ in terms of the ratio $n_2/n_1$. Also shown are the single component limit \eqref{e:singlemu} ($\times$)  and the impurity limit  \eqref{e:IA_Deltamu} ($\ocircle$). $\{a_{11},a_{22},a_{12}\}/a_0 = \{50,120,100\}$ with $\add_{ij}$ as in Fig.~\ref{f:LHY_integ}.}
\end{figure}

Before the calculation we numerically evaluate the $\theta_{k}$ integrals to construct a one-dimensional lookup table of $\Delta\mu_i/n_2^{3/2}$ for $n_1/n_2 \in [0,1]$ [e.g. see Fig.~\ref{f:f1D_lookup}(a)]. We also evaluate a similar lookup table but for the inverse ratio $n_2/n_1 \in [0,1]$ [e.g. see Fig.~\ref{f:f1D_lookup}(b)], noting that  the two tables cover the full range of possible density ratios: i.e.~to evaluate  Eq.~\eqref{e:Deltamu} for an inhomogeneous density we interpolate from the first lookup table for $\Delta\mu_i/n_2^{3/2}$ if $n_1/n_2<1$ or from the second for $\Delta\mu_i/n_1^{3/2}$ if $n_2/n_1<1$.
Note that the second lookup table is produced by evaluating  
\begin{equation}
    \frac{\Delta\mu_i}{n_1^{3/2}} = \frac{m^{3/2}}{3\sqrt2\pi^2\hbar^3}\sum_\pm\int_0^{\pi/2} d\theta_k\sin\theta_k \Re I^\downarrow_{i\pm}(n_2/n_1)\label{e:Deltamu_scaled2},
\end{equation}
using results similar to Eqs.~ (\ref{eq:Eup})-(\ref{e:QF_integrand_scaled}), but defined in terms of the ratio $n_2/n_1$.

\subsection{Stationary droplet states}\label{s:SDS}

We consider a  Dy mixture with dipole length $\add_{ij} = 130.1a_0$. 
To classify droplets as miscible or immiscible, we define the overlap \begin{align}
    \chi_{12} = \frac1{\sqrt{N_1N_2}}\int d\bx\, \psi_1^*\psi_2.
\end{align}
 When both components overlap perfectly (i.e.~are single-modal) $\chi_{12} = 1$ and the droplet is perfectly miscible, and when $\chi_{12}=0$ there is no overlap between the components and the droplet is perfectly immiscible (for stationary states of this system we take $\psi_i$ to be real and positive). 


\subsubsection{Immiscible droplets}

\begin{figure}
\includegraphics[width = 3.4in ]{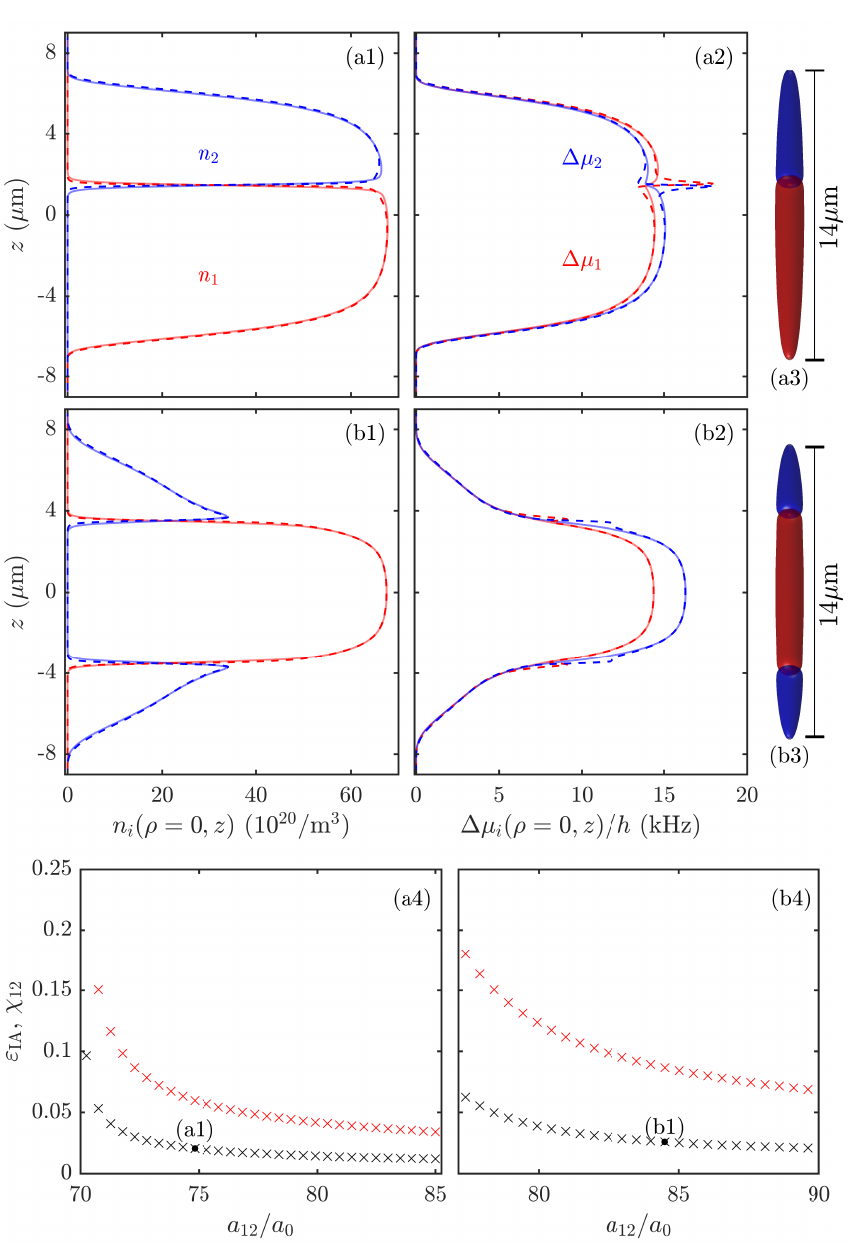}
\caption{\label{f:immisc_dense} Immiscible BMG droplets. (a1,b1) Axial density of component 1 (red) and 2 (blue), (a2,b2) corresponding chemical potential corrections, (a3,b3) isodensity surfaces at $n_i=\SI{2e21}{\per\metre\cubed}$ and (a4,b4) $\varepsilon_\text{IA}$ from \eqref{e:errorIA}  (red) and overlap EGPE states $\chi_{12}$ (black). Showing the EGPE (solid), and  IA (dashed). Parameters are $a_{11}=70a_0$, $N_1=10^4$, $N_2 = \num{5e3}$, and 
(a) $a_{22} = 70a_0$, $a_{12} = 75a_{0}$, 
(b) $a_{22} = 85a_0$, $a_{12} = 84.5a_{0}$.}
\end{figure}

In Fig.~\ref{f:immisc_dense} we show results for ground state droplets that are immiscible. This case is obtained by an appropriate choice of  $s$-wave interactions, with the atoms separating into two or three distinct domains. The domains separate axially, reducing the interface area, and hence decreasing the interface energy.

When $a_{11}\approx a_{22}$ there is no preference over which component is at the center, so the ground state has one domain of each component, giving an asymmetric droplet, e.g.~see Fig.~\ref{f:immisc_dense}(a3) for $a_{11}=a_{22}$  (also see Fig.~4 of \cite{Bisset2021a} for a case with a slightly imbalance of scattering lengths).
In contrast, when the intraspecies scattering lengths are sufficiently different, the advantage of having one component at the center and the other component at each end outweighs the cost of the additional domain boundary, giving a symmetric immiscible droplet [Fig.~\ref{f:immisc_dense}(b)].


Using the IA, Eq.~\eqref{EGPEIA}, we have computed stationary states and in Figs.~\ref{f:immisc_dense}(a1) and (b1) we compare the density profiles on the $z$-axis to those of  the full EGPE. In both cases there is excellent agreement between the two theories, albeit some discrepancy  near the domain boundaries where $n_1\approx n_2$ and the validity condition for the IA fails. In Figs.~\ref{f:immisc_dense}(a2) and (b2) we compare $\Delta\mu_i$ for these cases. This also shows good agreement between the theories within the domains but that the IA significantly overestimates the value of $\Delta\mu_i$  in the narrow interface region. 

We quantify the error between the EGPE and the IA theory using the measure
\begin{equation}
    \varepsilon_\text{IA} = \frac1N \sum_{i}\int d\bx |n_i(\bx) - n_i^{\text{IA}}(\bx)|,\label{e:errorIA}
\end{equation}
where  $n_i^\text{IA}$ is the density of component $i$ calculated using the IA. In Figs.~\ref{f:immisc_dense}(a4) and (b4) we plot Eq.~\eqref{e:errorIA} for the two different cases, and the overlap $\chi_{12}$. Here the overall error is small when deep in the immiscible region and increases as the immiscibility threshold is reached. Our results show that the error increases with increasing $\chi_{12}$. This occurs because $\chi_{12}$ is related to the amount of overlap between the two components and in the immiscible regime this is dominated by the interfaces where the IA is poor. For example, in Figs.~\ref{f:immisc_dense}(a4) and (b4) we see that  $\varepsilon_\text{IA}$ is larger for the symmetric immiscible case  of Fig.~\ref{f:immisc_dense}(b3)   than the  asymmetric case of Fig.~\ref{f:immisc_dense}(a3). This is due to the additional interface in the symmetric case.

\subsubsection{Miscible droplets} 

\begin{figure}
    \includegraphics[width = 3.4in]{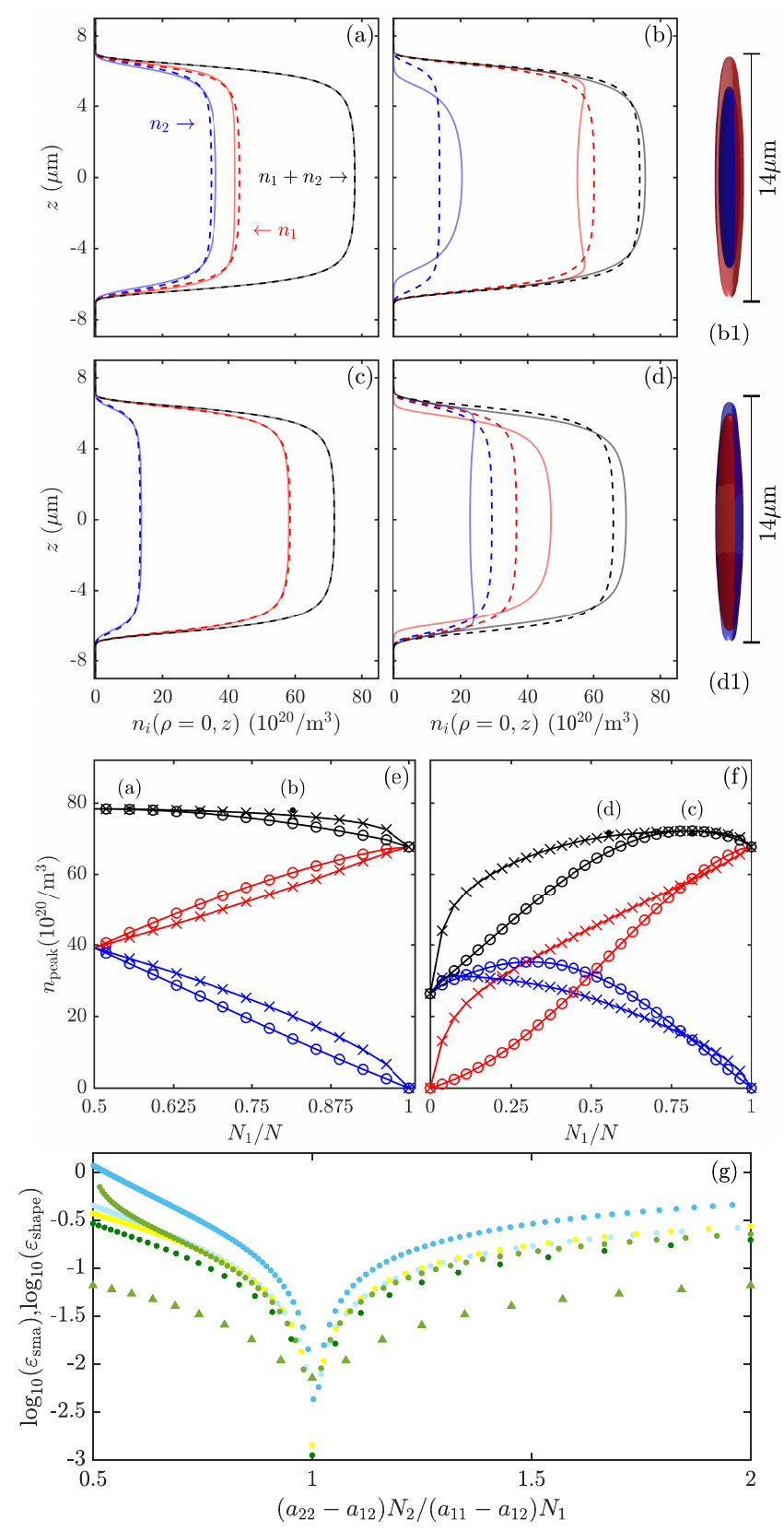}
\caption{\label{f:misc_axial_density} Miscible BMG droplets. (a-d) axial density profiles of component 1 (red), component 2 (blue) and total density (black), using the EGPE (solid), and SMA (dashed). (b1) and (d1) show the $n_i=\SI{7.5e20}{\per\metre\cubed}$ isodensity surfaces for (b) and (d) respectively. 
(e,f) Comparison of total (black), component 1 (red), and component 2 (blue) peak densities as $N_1/N$ varies, using the EGPE ($\times$) and SMA ($\ocircle$). Scattering lengths are (a-f) $a_{11}=70a_0$, $a_{12} = 65a_0$, (a,b,e) $a_{22}=70a_0$, and (c,d,f) $a_{22}=85a_0$. 
The total number of atoms in (a-f) $N=\num{2.7e4}$, with (a,d) $N_1 = \num{1.5e4}$,  (b,c) $N_1 = \num{2.2e4}$.
(g) plot of  $\varepsilon_\mathrm{sma}$ from \eqref{e:errorSMA} ($\blacktriangle$) and  $\varepsilon_\mathrm{shape}$ from \eqref{e:errorShape} ($\bullet$) with $N_{\mathrm{total}} = 27\times 10 ^{3}$, except for (yellow) for which $N_{\mathrm{total}} = 5\times 10 ^{3}$, and \{$a_{11}$, $a_{22}$, $a_{12}\}/a_0$  
(blue): \{$85$, $70$, $30$\}, 
(light-blue): \{$85$, $70$, $60$\},
(yellow): \{$85$, $70$, $65$\},
(light-green): \{$70$, $70$, $65$\},
(green): \{$85$, $70$, $65$\}.
}
\end{figure}
 
In Fig.~\ref{f:misc_axial_density} we have chosen the $s$-wave interactions so that droplet is miscible. In Figs.~\ref{f:misc_axial_density}(a)-(f) the total number of atoms has been fixed such that $N= \num{2.7e4}$, but vary the relative numbers in the components. We show isodensity surfaces of droplet states for $N_1/N \approx 0.81$ in Fig.~\ref{f:misc_axial_density}(b1) and for $N_1/N \approx 0.56$ in Fig.~\ref{f:misc_axial_density}(d1).

We compare the $z$-axis density profiles of stationary droplet states obtained from the EGPE and the SMA in Figs.~\ref{f:misc_axial_density}(a)-(d).  When the intraspecies scattering lengths are the same and the number imbalance is small [see Fig.~\ref{f:misc_axial_density}(a) where $N_1/N \approx 0.56$], the two components have similar spatial variation i.e.~are close to being single modal (in the sense of Sec.~\ref{sec:singlemodal}). Here  the SMA is a good approximation for the individual densities, and provides an even  better approximation to the total density profile. 

The SMA is generally a poor approximation when there is a large number imbalance [Fig.~\ref{f:misc_axial_density}(b)] or a large difference between the intraspecies scattering lengths [Fig.~\ref{f:misc_axial_density}(d)]. In these cases the component with the larger number or larger intraspecies scattering length can lower its energy by lowering  its peak density and increasing its width, i.e.~favouring the components taking different shapes.

However, differences in number and intraspecies scattering lengths can compensate and allow the SMA to work well in imbalanced situations. For example, in Fig.~\ref{f:misc_axial_density}(c) we consider a case where  $N_1\gg N_2$ and $a_{22}>a_{11}$.  
We can quantify the single-modality of an EGPE solution by the measure
\begin{equation}
    \varepsilon_\mathrm{shape} = \int d\bx \left|\frac{n_1(\bx)}{N_1} - \frac{n_2(\bx)}{N_2}\right|,\label{e:errorShape}
\end{equation}
such that $ \varepsilon_\mathrm{shape}=0$ if the component wavefunctions are single-modal. Additionally we can adapt   \eqref{e:errorIA} to measure the difference between the SMA and the EGPE solution as
 \begin{equation}
    \varepsilon_\text{sma} = \frac1N \sum_{i}\int d\bx |n_i(\bx) - n_i^{\text{sma}}(\bx)|,\label{e:errorSMA}
\end{equation}
where $n_i^{\text{sma}}$ is the $i$-component density obtained from the SMA.
In Fig.~\ref{f:misc_axial_density}(g) we use the $\varepsilon_\mathrm{shape}$  and $\varepsilon_\text{sma}$ to characterise the single-modality of the ground state droplets, and the error of the SMA approximation for droplets with a range of interaction parameters and atom numbers. Importantly these results show that the closer the parameters are to satisfying Eq.~(\ref{smaCrit}) then the more single-modal the EGPE solution is and the SMA is more accurate.
%

In Fig.~\ref{f:misc_axial_density}(e)  we plot the peak droplet densities obtained from the EGPE and the SMA  as  $N_1$ varies for $a_{11}=a_{22}$. These results show that SMA is identical to the EGPE result  for $N_1\to 0$, $N/2$, or $N$ and results are symmetric about $N_1=N/2$ (recalling that $N_2=N-N_1$).
In Fig.~\ref{f:misc_axial_density}(f), we show corresponding results for $a_{11}\ne a_{22}$, showing poor agreement for $N_1<N/2$.

\begin{figure*}
    \includegraphics[height = 4.5in ]{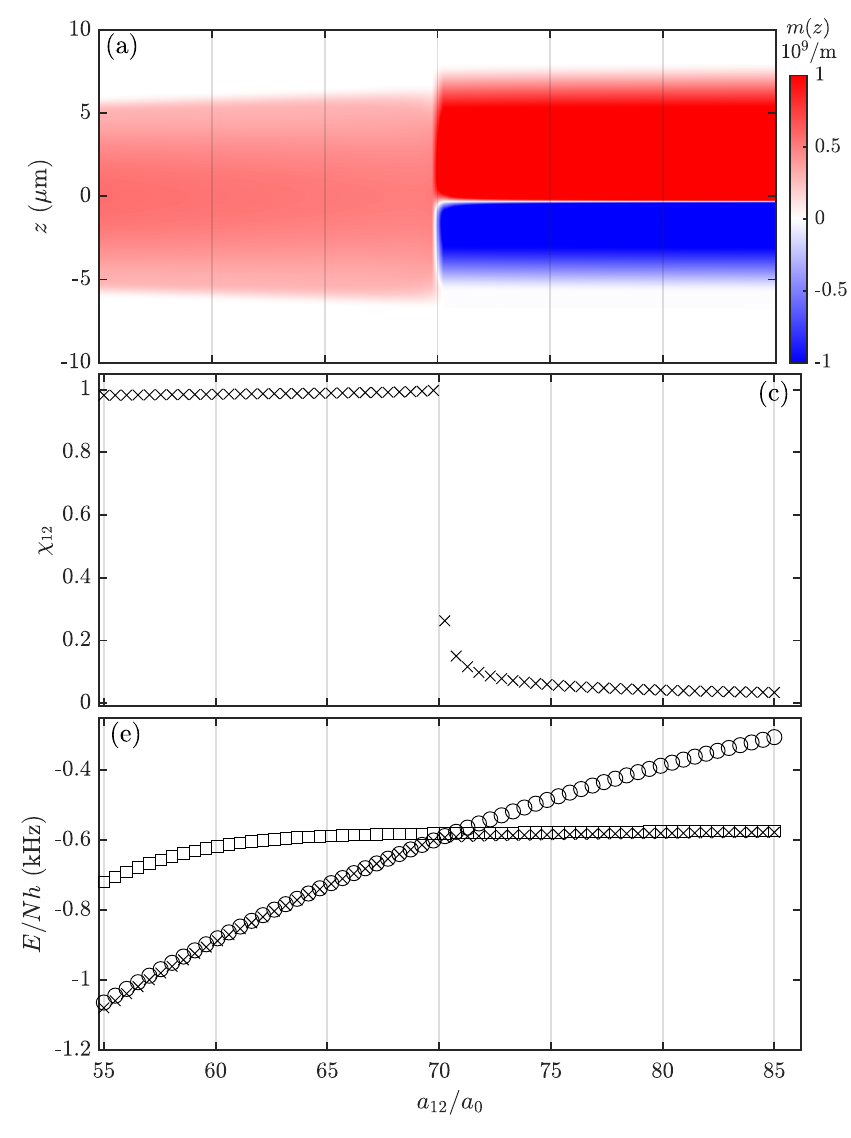} 
    \includegraphics[trim=10 0 0 0,clip = true,height = 4.5in ]{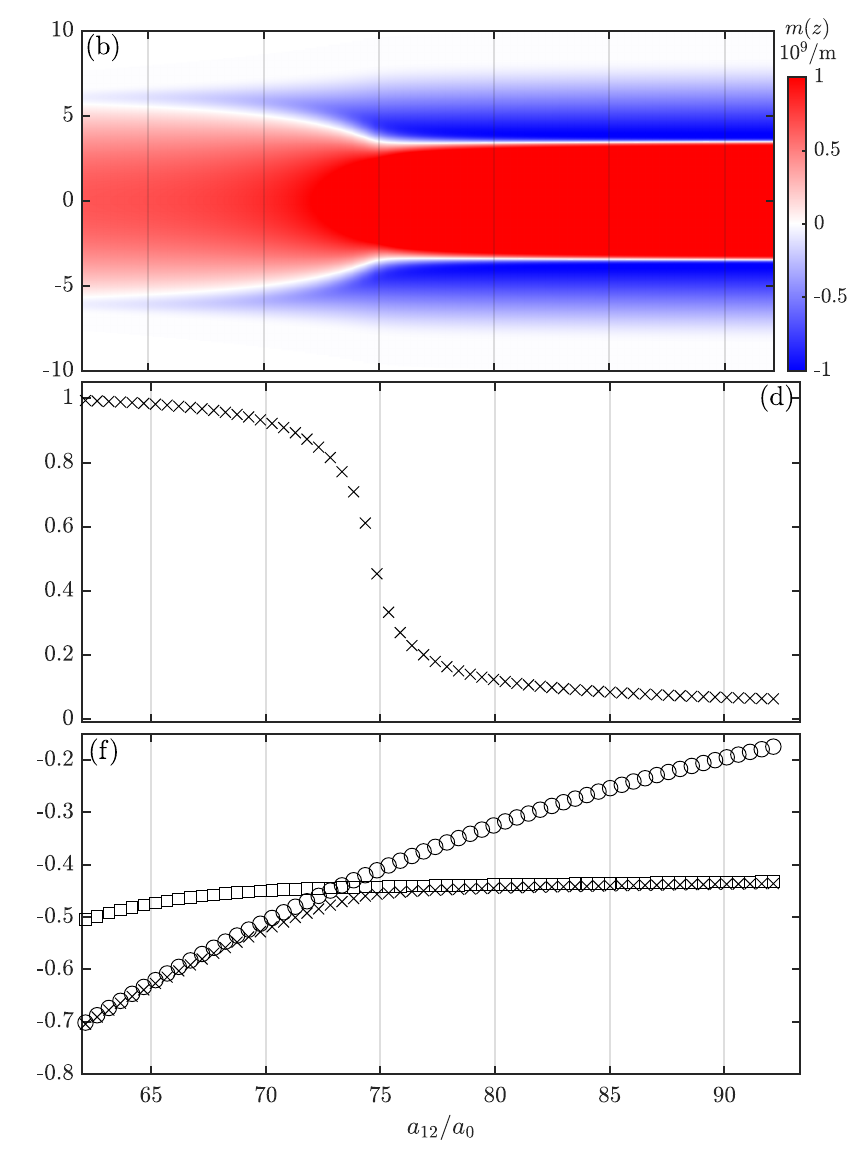}
\caption{\label{f:spin_dense} 
(a,c,e) Miscible to asymmetric immiscible transition with 
$a_{11}=a_{22}=70a_0$, $N_1=10^4$, and $N_2=\num{5e3}$. 
(b,d,f) Miscible to symmetric immiscible transition with  
$a_{11}=70a_0$, $a_{22}=85a_0$, $N_1=10^{4}$, and $N_2=\num{5e3}$. 
(a,b) pseudo-spin density $m_z$, (c,d) component overlap $\chi_{12}$, and (e,f) the energy per particle $E/Nh$, as $a_{12}$ varies. In (e,f) we compare EGPE ($\times$) with the SMA ($\ocircle$) and IA ($\square$). 
 }
\end{figure*} 

\subsection{Miscible-immiscible transition}\label{s:imt}

In Fig.~\ref{f:spin_dense} we have solved the EGPE for the ground state droplets as $a_{12}$ varies. We start from a state which is deep in the miscible regime (i.e.~with $a_{12}\ll\sqrt{a_{11}a_{22}}$) and increase $a_{12}$ until we are deep in the immiscible regime ($a_{12}\gg\sqrt{a_{11}a_{22}}$).
To characterize the density domain structure that emerges as a result of the miscible-immiscible transition, we introduce the linear pseudo-spin density 
\begin{align}
m_z(z) = \int dxdy \,(|\psi_1|^2 - |\psi_2|^2).
\end{align}
For the results shown in Fig.~\ref{f:spin_dense}(a) we see there is little variation in $m_z$ when the droplet is miscible (note that  $m_z>0$ here because $N_1>N_2$), 
and $\chi_{12}$ is maximized with a value close to unity [see Fig.~\ref{f:spin_dense}(c)].  
When the miscible-immiscible transition is crossed (at $a_{12}\approx70\,a_0$), the ground state sharply changes to an immiscible droplet with two domains. This is signaled by the emergence of both red ($m_z>0$) and blue ($m_z<0$) domains in Fig.~\ref{f:spin_dense}(a), as well as the discontinuous change in $\chi_{12}$. After the  transition $\chi_{12}$ decreases with increasing $a_{12}$ as the interface between the two components becomes narrower.
In Fig.~\ref{f:spin_dense}(e) we compare the energy predicted by the SMA and the IA to the groundstate energy of the EGPE. There is good agreement between the EGPE and SMA energies when the droplets are miscible, and good agreement between the EGPE and the IA when the droplets are immiscible. The energies predicted by the SMA and the IA intersect approximately at the miscible-immiscible phase transition. 

\begin{figure*}
    \includegraphics[trim=160 40 50 200,clip=true,height = 4.5in ]{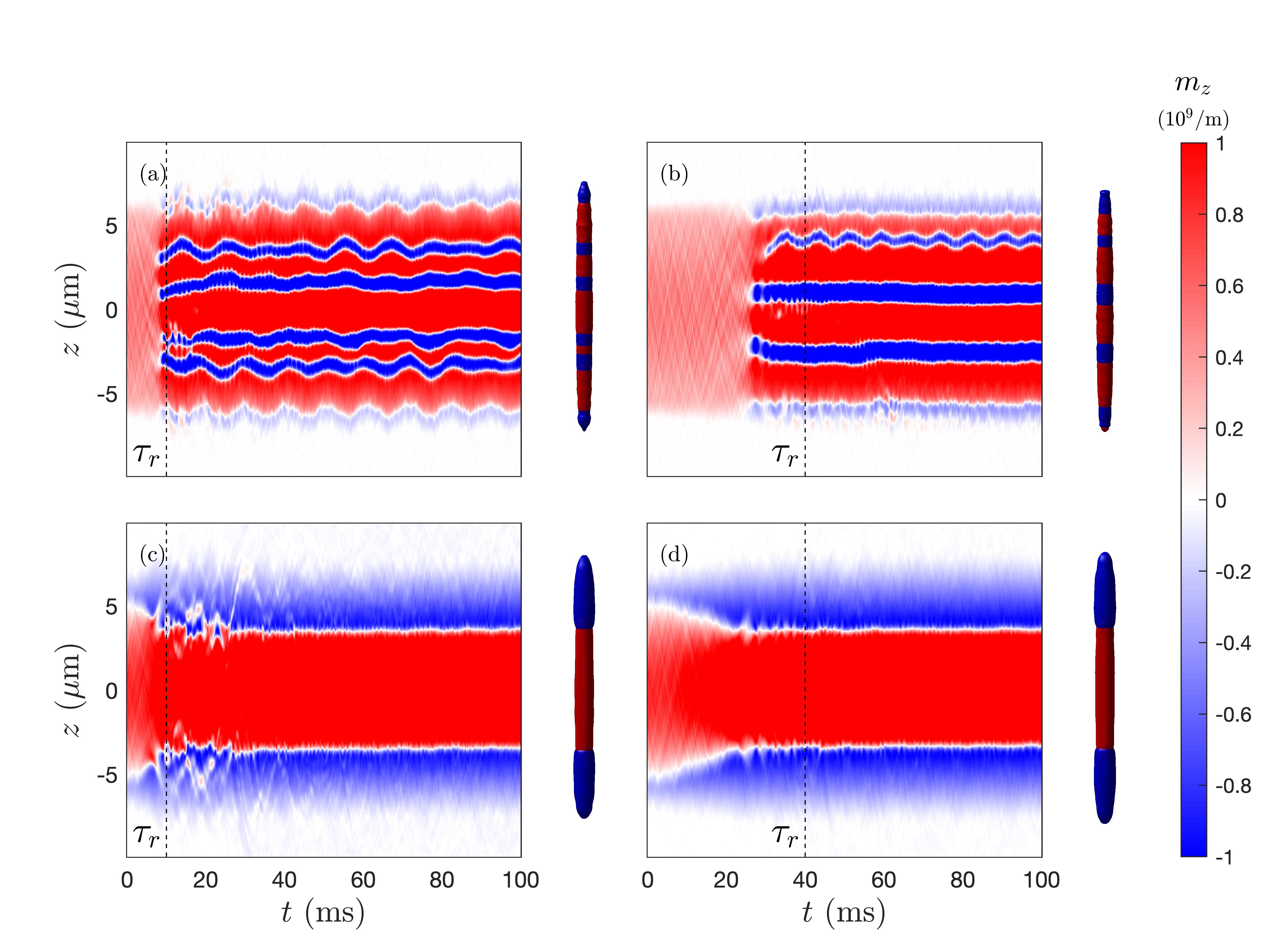}
\caption{\label{f:dynamics} Preparation of immiscible droplets from an initial miscible droplet.
(a,b) $a_{12}$ from 65 to $75a_0$ with other parameters as in Fig.~\ref{f:spin_dense}(a,c,e),
(c,d) $a_{12}$ from 70 to $80a_0$ with other parameters as in Fig.~\ref{f:spin_dense}(b,d,f),
(a,c) using a $\tau_r = 10$ms and (b,d) a 40ms ramp.} 
\end{figure*}

In Figs.~\ref{f:spin_dense}(b), (d) and (f) we consider a case with $a_{11}\neq a_{22}$. Both the spin-density [Fig.~\ref{f:spin_dense}(b)] and the overlap [Fig.~\ref{f:spin_dense}(d)] are notably different from the $a_{11}=a_{22}$ case. When $a_{12}<\sqrt{a_{11}a_{22}}$  the pseudo-spin density $m_z$ already changes sign near the edges of the droplet (i.e.~faint blue domain). In this regime the profiles are strongly overlapped (miscible), but the component density profiles have a different shape [e.g.~the component profiles for this case at $a_{12}=65\,a_0$ are shown in Fig.~\ref{f:misc_axial_density}(d)].
As $a_{12}$ increases the magnitude of $m_z$ increases and the overlap $\chi_{12}$ decreases gradually indicating a smooth transition into the immiscible state  [e.g.~the component profiles for this case at $a_{12}=84.5\,a_0$ are shown in Fig.~\ref{f:immisc_dense}(b1)].
We compare the predicted energies of the EGPE, IA, and SMA in Fig.~\ref{f:spin_dense}(f), with good agreement between the approximate theories and the EGPE in their regimes of validity. 

\subsection{Dynamic simulations of the miscible-immiscible transition}\label{s:dyn}

We explore the real-time dynamics of the system with a ramp from the miscible to the immiscible phase. We evolve the time-independent EGPE $i\hbar\dot{\psi}_i = \ML_i\psi_i$ using a fourth order Runge-Kutta method maintaining cylindrical symmetry\footnote{Full three-dimensional calculations \cite{Smith2021a} demonstrate that domains  form axially through the immiscibility transition, allowing us to perform dynamical simulations assuming cylindrical symmetry.}, with $\ML_i$ given by \eqref{e:GP_operator}. A small amount of noise is added the initial state ground state droplet solution to mimic the effects of quantum and thermal noise.
We then perform a linear ramp increasing of the value of $a_{12}$ over a time period $\tau_r$ after which the $a_{12}$ is held constant and the state is allowed to evolve in time. 

In Fig.~\ref{f:dynamics} we consider simulations of ramps for the two cases presented stationary states for in Sec.~\ref{s:imt} and for two different ramp times:  $\tau_r=10$ms in Figs.~\ref{f:dynamics}(a) and (c), and $\tau_r = 40$ms in Fig.~\ref{f:dynamics}(b) and (d). 
 Results vary from shot to shot due to statistical variation of the noise, but the results shown here are typical. We find both sets of parameters produce a dynamically stable droplets which are long lived (noting that we do not allow for three-body loss). For shorter ramp times, we find that collective breathing-modes are more pronounced than they are in the slower ramps. These modes are revealed by the oscillating axial extent of the droplets in the simulations.

When $a_{11}=a_{22}$ [Fig.~\ref{f:dynamics}(a) and (b)], the miscible-immiscible transition produces a droplet that is different from the ground state [cf.~Fig.~\ref{f:spin_dense}(a)], notably several small domains develop. More domains are produced with a fast ramp than in a slow ramp. These results are consistent with domains arising as defects when the system is ramped across a discontinuous phase transition [cf.~the discontinuous change in the overlap of the states $\chi_{12}$ in Fig.~\ref{f:spin_dense}(c)].

We find very different dynamics when $a_{11}\neq a_{22}$ [see Fig.~\ref{f:dynamics}(c) and (d)]. In this case the droplet is able to dynamically follow states which are similar to the ground states seen in Fig.~\ref{f:spin_dense}(b), as the overlap $\chi_{12}$ varies smoothly [Fig.~\ref{f:spin_dense}(d)]. For faster ramps we do not see the emergence of more domains. Larger domain wall oscillations are present in faster ramps, but these do not persist on long time scales. 

\section{Conclusions}\label{s:con}

Full EGPE calculations of quantum droplet states of a BMG are numerically intensive and evaluating the quantum fluctuation term involves numerical integration. Here we have developed two useful approximations that lead to a simpler calculations and insight into the role of the quantum fluctuation term in the miscible and immiscible regimes. The SMA generalises an idea widely used for binary (non-magnetic) droplets to the magnetic case, notably this describes miscible droplets but without making the assumption of density locking that is usually employed in the non-magnetic case. The IA provides a description of the immiscible regime of the BMG droplet. For this case we have developed analytic results for the quantum fluctuation term. We have also outlined important aspects of the numerical solution of the EGPE and the approximate theories, and described a novel procedure to accurately evaluate the quantum fluctuation term of the EGPE using a one-dimensional lookup table parameterized by the density ratio of the components.   We have compared the SMA and IA to results of the EGPE over a wide range of parameters, showing excellent agreement in their regimes of validity.
 
We present results showing the ground state droplets crossing the transition to immiscibility by varying the interspecies contact interaction, showing that the transition can be smooth or discontinuous, depending on the parameter regime. In this study we also see that the SMA and IA work well for the miscible and immiscible regimes, respectively. Also, by comparing the energy of the two approximations, we can predict where the transition occurs. Finally, we have presented some dynamical simulations of a BMG droplet taken into the immiscible regime by a finite speed ramp of the interspecies contact interactions. For the continuous transition, the droplet is able to resemble the ground states across the ramp, while for the discontinuous case, multiple small domains are observed to develop.

 Our work opens the door for more studies of BMG droplets and will hopefully support future experimental work to produce and understand this new class of quantum droplet. An important aspect will be to account for three-body loss, which is usually the dominant loss mechanism and will set the time-scale for experiments to prepare and manipulate BMG droplets.


%

\end{document}